\documentclass[12pt]{article}
\begin{document}
\title{The nonlinear viscoelastic behavior of polypropylene}

\author{Aleksey D. Drozdov\footnote{Corresponding author} and Jesper deClaville Christiansen\\
Department of Production\\
Aalborg University\\
Fibigerstraede 16\\
DK--9220 Aalborg, Denmark}
\date{}
\maketitle

\begin{abstract}
A series of tensile relaxation tests is performed 
on isotactic polypropylene in the sub-yield and post-yield regions
at room temperature.
Constitutive equations are derived for the time-dependent
response of a semicrystalline polymer at isothermal loading
with small strains.
Adjustable parameters in the stress--strain relations are found
by fitting experimental data.
It is demonstrated that the growth of the longitudinal strain 
results in an increase in the relaxation rate in a small interval 
of strains in the sub-yield domain.
When the strain exceeds some critical value 
which is substantially less than the apparent yield strain,
the relaxation process becomes strain-independent.
\end{abstract}
\newpage

\section{Introduction}

This paper is concerned with the nonlinear viscoelastic behavior
of semicrystalline polymers at isothermal uniaxial deformation 
with small strains.
The study concentrates on the time-dependent response of isotactic 
polypropylene at room temperature, 
but we believe that the constitutive equations to be
derived may be applied to other semicrystalline polymers 
[polyamides, polyethylene, polytetrafluorethylene,
poly(ethylene terephthalate),
poly(butylene terephthalate),
poly(ethylene-2,6-naphthalene dicarboxilate), etc.]
at temperatures above the glass transition point for the amorphous 
phase.
The viscoelastic and viscoplastic behavior of polypropylene (PP)
has been a focus of attention in the past decade, which
may be explained by numerous applications of this material in
industry (oriented films for packaging,
reinforcing fibres,
nonwoven fabrics,
polyethylene--polypropylene co\-po\-lymers,
blends of PP with thermoplastic elastomers, etc.).

The nonlinear viscoelastic response of polypropylene was studied
by Ward and Wolfe \cite{WW66}, see also \cite{WH93},
and Smart and Williams \cite{SW72} three decades ago,
and, more recently, by Ariyama \cite{Ari93a,Ari93b,Ari96,AMK97},
Wortmann and Schulz \cite{WS94,WS95},
Dutta and Edward \cite{DE97},
and Read and Tomlins \cite{RT97}.
In the past couple of years, the linear viscoelastic behavior
of PP fibres was analyzed by Andreassen \cite{And99}
and Lopez-Manchado and Arroyo \cite{LA00},
and the dynamic viscoelastic response of blends of isotactic polypropylene
with polyethylene and styrene-butadiene-styrene triblock copolymer 
was studied by Souza and Demarquette \cite{SD02} and 
Gallego Ferrer et al. \cite{GSV00}, respectively.
It was found that the loss tangent of isotactic polypropylene 
demonstrates two pronounced maxima being plotted versus
temperature.
The first maximum ($\beta$--transition in the interval 
between $T=-20$ and $T=10$ $^{\circ}$C) is associated with 
the glass transition in the most mobile part of the amorphous phase, 
whereas the other maximum ($\alpha$--transition in the interval 
between $T=50$ and $T=80$ $^{\circ}$C) is attributed to the
glass transition in the remaining part of the amorphous phase.
This conclusion is confirmed by DSC (differential scanning
calorimetry) traces for quenched PP that reveal an endoterm at $T=70$ 
$^{\circ}$C which can be ascribed to thermal activation of  
amorphous regions with restricted mobility under heating \cite{SSE99}.

Scanning electron microscopy and atomic force microscopy 
evidence that isotactic PP is a semicrystalline polymer 
with a complicated morphology. 
The crystalline phase contains brittle monoclinic $\alpha$ 
spherulites (the characteristic size of 100 $\mu$m)
consisting of crystalline lamellae (10 to 20 nm in thickness)
directed in radial and tangential directions \cite{CCG98},
and ductile hexagonal $\beta$ spherulites consisting
of parallel-stacked lamellae \cite{LFG99}.
The amorphous phase is located  between spherulites
and inside the spherulites between lamellae.
It consists of (i) relatively mobile chains between spherulites
and between radial lamellae and (ii) severely restricted
chains in regions bounded by radial and tangential lamellae.
Mechanical loading results in inter-lamellar separation,
rotation and twist of lamellae,
fine and coarse slip of lamellar blocks,
and their fragmentation.
The latter leads to reorganization of blocks and strain-induced
smectic--monoclinic transitions \cite{SSE99}.
Straining of polypropylene causes chain slip through 
the crystals, breakage and reformation of tie chains,
and activation of restricted amorphous regions driven
by lamellar disintegration.
In the post-yield region, these transformations of microstructure
lead to the onset of voids between lamellae,
break-up of crystalls, and formation of fibrills \cite{ZBC99}.

It is hard to believe that these changes in the internal structure
of PP may be adequately described by a constitutive model 
with a relatively small number of adjustable parameters.
To develop stress--strain relations for polypropylene,
we apply the method of ``homogenization of micro-structure"
\cite{BKR02}.
According to this approach, an individual (equivalent) phase 
is taken into account whose deformation captures 
essential features of the response of a semicrystalline 
polymer with a sophisticated morphology.

An amorphous phase is chosen as the equivalent phase
because of the following reasons:
\begin{enumerate}
\item
The viscoelastic response of semicrystalline polymers 
is conventionally associated with rearrangement of chains 
in amorphous regions \cite{CCG98}.

\item
According to the Takayanagi--Nitta concept \cite{NT99,NT00},
sliding of tie chains along and their detachment from lamellae 
play the key role in the time-dependent response of
semicrystalline polymers.

\item
The viscoplastic flow in semicrystalline polymers
is assumed to be ``initiated in the amorphous phase 
before transitioning into the crystalline phase" \cite{MP01},

\item
Conventional models for  polyethylene \cite{BKR02},
polypropylene  \cite{SW95,SW96}
and poly(ethylene terephthalate) \cite{LB99,BSL00}
treat these polymers as networks of macromolecules.
\end{enumerate}

Above the glass transition temperature for the mobile 
amorphous phase, polypropylene is thought of as a network 
of chains connected by junctions.
Deformation of a specimen induces slip of junctions 
with respect to the bulk material.
Sliding of junctions reflects slippage of tie molecules
along lamellae and fine slip of lamellar blocks
which are associated with the viscoplastic behavior
of a semicrystalline polymer.

The viscoelastic response of PP
is described in terms of the concent of transient networks
\cite{GT46,Yam56,Lod68,TE92},
and it is modelled as separation of active chains from 
their junctions and attachment of dangling chains to 
temporary nodes.
Unlike previous studies, the network of polymeric chains 
is assumed to be strongly inhomogeneous,
and it is treated as an ensemble of meso-regions (MR) with various 
potential energies for the detachment of active strands.
Two types of MRs are distinguished: (i) active domains where
strands separate from junctions as they are thermally agitated
(these MRs model a mobile part of the amorphous phase), 
and (ii) passive domains where detachment of
chains from junctions is prevented (these MRs reflect a part 
of the amorphous phase whose mobility is severely restricted 
by radial and tangential lamellae).
Detachment of active chains from the network is treated 
as a thermally-activated process whose rate is given by
the Eyring formula \cite{KE75} with a strain-dependent attempt rate.
Deformation of a specimen results in (i) an increase in the concentration
of active MRs (which is ascribed to a partial release of the amorphous phase
in passive meso-domains driven by fragmentation of lamellae)
and (ii) a mechanically-induced acceleration of separation 
of strands from their junctions in active MRs.

An increase in the rate of relaxation of amorphous glassy polymers 
at straining is conventionally assumed to occur in the entire region
of nonlinear viscoelasticity (from the strain $\epsilon$ of at least 0.005 
to the yield strain $\epsilon_{\rm y}$), see \cite{MCP92}.
It is ascribed to the mechanically-induced growth of ``free volume" 
between macromolecules which results in an increase in their
mobility \cite{KE87,LK92,OK95,CKP95}.
Despite wide applications of the free volume concept in nonlinear
viscoelasticity of solid polymers, experimental verification of this
approach was limited to relatively small strains (substantially below
the yield point).
The objectives of this study are 
(i) to report experimental data in tensile relaxation tests on 
isotactic polypropylene in the range of strains between 
$\epsilon=0.005$ (which corresponds to the end of
the region of linear viscoelasticity) and 
$\epsilon=0.16$ (at higher strains necking of specimens 
is observed at the chosen cross-head speed),
(ii) to derive constitutive equations for the time-dependent
behavior of semicrystalline polymers at isothermal uniaxial 
stretching with small strains,
(iii) to find adjustable parameters in the stress--strain relations
by fitting observations,
and (iv) to assess the interval of strains where the rate of
relaxation monotonically increases with strain intensity.

The exposition is organized as follows.
Section 2 is concerned with the description of the experimental
procedure.
Kinetic equations for sliding of junctions and detachment of active strands 
and attachment of dangling strands in active MRs
are developed in Section 3.
The strain energy density of a semicrystalline polymer 
is determined is Section 4.
In Section 5 stress--strain relations are derived for uniaxial deformation
of a specimen by using the laws of thermodynamics.
These constitutive equations are employed to fit experimental data
in tensile relaxation tests in Section 6.
Some concluding remarks are formulated in Section 7.

\section{Experimental procedure}

A series of uniaxial tensile relaxation tests on polypropylene specimens 
was performed at room temperature.
Isotactic polypropylene (Novolen 1100L) was supplied by Targor (BASF).
ASTM dumbbell specimens with length 14.8 mm, width 10 mm 
and height 3.8 mm were injection molded and used without 
any thermal pre-treatment.
Our DSC measurements (the sample mass 16.43 mg, 
the heating rate 10 K/min) 
demonstrated the specific enthalpy of melting 
$\Delta H_{\rm m}=95.5$ J/g, which corresponds to 
the degree of crystallinity $\kappa_{\rm c}=0.46$ 
(with reference to \cite{Wun80}, the enthalpy of fusion for a fully 
crystalline polypropylene is assumed to be 209 J/g).

Mechanical experiments were performed with a testing machine 
Instron--5568 equipped with electro-mechanical sensors 
for the control of longitudinal strains in the active zone of samples 
(the distance between clips was about 50 mm).
The tensile force was measured by the standard loading cell.
The engineering stress $\sigma$ was determined
as the ratio of the axial force to the cross-sectional area
of the specimens in the stress-free state (38 mm$^{2}$).

16 relaxation tests were carried out at longitudinal strains
in the range from $\epsilon_{1}=0.005$ to $\epsilon_{16}=0.16$,
which correspond to the domain of nonlinear viscoelasticity,
sub-yield and post-yield regions for isotactic polypropylene 
(the yield strain, $\epsilon_{\rm y}$, 
is estimated by the supplier as 0.13).

Any relaxation test was performed on a new sample.
No necking of specimens was observed in experiments.
In the $k$th relaxation test ($k=1,\ldots,16$), a specimen was loaded 
with the constant cross-head speed 5 mm/min (which roughly 
corresponded to the strain rate  $\dot{\epsilon}_{0}=0.05$ min$^{-1}$) 
up to the longitudinal strain $\epsilon_{k}$, which was preserved
constant during the relaxation time $t_{\rm r}=20$ min.
The engineering stresses, $\sigma$, at the beginnings of relaxation
tests are plotted in Figure 1 together with the stress--strain curve for
the specimen strained up to $\epsilon_{16}=0.16$.
The figure demonstrates fair repeatability of experimental data.

The longitudinal stress, $\sigma$, is plotted versus  the logarithm 
($\log=\log_{10}$) of time $t$ (the initial instant $t=0$ corresponds 
to the beginning of the relaxation process) in Figures 2 to 4.
These figures show that the stress, $\sigma$, monotonically increases
with strain, $\epsilon$, up to $\epsilon_{0}=0.08$
and remains practically constant when the strain exceeds 
the critical strain $\epsilon_{0}$.

\section{A micro-mechanical model}
 
A semicrystalline polymer is treated as a temporary network of 
chains bridged by junctions.
The network is modelled as an ensemble of meso-regions with
various strengths of interaction between macromolecules.
Two types of meso-domains are distinguished: passive and active.
In passive MRs restricted by radial and tangential lamellae, 
inter-chain interaction prevents detachment of 
chains from junctions, which implies that all  nodes in these domains
are thought of as permanent.
In active MRs, active strands (whose ends are connected to 
contiguous junctions) separate from temporary junctions 
at random times when these strands are thermally agitated.
An active chain whose end slips from a junction is transformed
into a dangling chain.
A dangling chain  returns into the active state
when its free end captures a nearby junction at a random instant.

Denote by $X$ the average number of active strands per unit mass
of a polymer.
Let $X_{\rm a}$ be the number of strands merged with the network 
in active MRs, and $X_{\rm p}$ the number of strands connected 
to the network in passive MRs.
Under stretching some crystalline lamellae (that restrict mobility
of chains in passive MRs) break,
which results in the growth of the number of strands to be rearranged.
As a consequence, the number of strands in active MRs increases
and the number of strands in passive meso-domains decreases.
This implies that the quantities $X_{\rm a}$ and $X_{\rm p}$ 
become functions of the current strain, $\epsilon$, which
obey the conservation law
\begin{equation}
X_{\rm a}(\epsilon) +X_{\rm p}(\epsilon) =X.
\end{equation}
Separation of active strands from the network and reformation of
dangling strands in active MRs are thought of as
thermally activated processes.
It is assumed that detachment of active strands from their junctions
is governed by the Eyring equation \cite{KE75}, where different 
meso-domains are characterized by different activation energies, 
$\bar{\omega}$, for separation of active strands from the network.

According to the theory of thermally-activated processes, 
the rate of separation of active strands in a MR with potential energy 
$\bar{\omega}$ in the stress-free state is given by
\[
\Gamma=\Gamma_{\rm a}\exp\biggl (-\frac{\bar{\omega}}{k_{\rm B}T}\biggr ),
\]
where $k_{\rm B}$ is Boltzmann's constant, 
$T$ is the absolute temperature, 
and the pre-factor $\Gamma_{\rm a}$ is independent
of energy $\bar{\omega}$ and temperature $T$.
Introducing the dimensionless potential energy
\[  \omega=\frac{\bar{\omega}}{k_{\rm B}T_{0}}, \]
where $T_{0}$ is some reference temperature,
and disregarding the effects of small increments of
temperature, $\Delta T=T-T_{0}$, on the rate of detachment, $\Gamma$,
we arrive at the formula
\begin{equation}
\Gamma=\Gamma_{\rm a}\exp (-\omega).
\end{equation}
It is assumed that Eq. (2) is satisfied for an arbitrary
loading process, provided that the attempt rate, $\Gamma_{\rm a}$,
is a function of the current strain,
\[ \Gamma_{\rm a}=\Gamma_{\rm a}(\epsilon). \]
The distribution of active MRs with various potential 
energies is described by the probability density $p(\omega)$ that equals the
ratio of the number, $N_{\rm a}(\epsilon,\omega)$, of active meso-domains 
with energy $\omega$ to the total number of active MRs,
\begin{equation}
N_{\rm a}(\epsilon,\omega)=X_{\rm a}(\epsilon)p(\omega).
\end{equation}
We suppose that the distribution function for potential energies 
of active MRs, $p(\omega)$, is strain-independent.

The ensemble of active meso-domains is described by the
function $n_{\rm a}(t,\tau,\omega)$ that equals the number of
active strands at time $t$ (per unit mass) belonging to
active MRs with potential energy $\omega$ that have last been 
rearranged before instant $\tau\in [0,t]$.
In particular, $n_{\rm a}(0,0,\omega)$ is the number (per unit mass)
of active strands in active MRs with potential 
energy $\omega$ in a stress-free medium,
\begin{equation}
n_{\rm a}(0,0,\omega)=N_{\rm a}(0,\omega), 
\end{equation}
and $n_{\rm a}(t,t,\omega)$ is the number (per unit mass)
of active strands in active MRs with potential 
energy $\omega$ in the deformed medium at time $t$
(the initial time $t=0$ corresponds to the instant when
external loads are applied to a specimen),
\begin{equation}
n_{\rm a}(t,t,\omega)=N_{\rm a}(\epsilon(t),\omega). 
\end{equation}
The amount
\[ \frac{\partial n_{\rm a}}{\partial \tau}(t,\tau,\omega)\biggl |_{t=\tau} d\tau \]
equals the number (per unit mass) of dangling strands in 
active MRs with potential energy $\omega$  that merge 
with the network within the interval $[\tau,\tau+d\tau ]$,
and the quantity
\[ \frac{\partial n_{\rm a}}{\partial \tau}(t,\tau,\omega) d\tau \]
is the number of these strands that have not detached from
temporary junctions during the interval $[\tau, t]$.
The number (per unit mass) of strands in active MRs that separate (for
the first time) from the network within the interval $[t,t+dt]$ reads
\[ -\frac{\partial n_{\rm a}}{\partial t}(t,0,\omega) dt, \]
whereas the number (per unit mass) of strands in active MRs 
that merged with the network during the interval $[\tau,\tau+d\tau ]$
and, afterwards, separate from the network
within the interval $[t,t+dt]$ is given by
\[ -\frac{\partial^{2} n_{\rm a}}{\partial t\partial \tau}(t,\tau,\omega) dt d\tau. \]
The rate of detachment, $\Gamma$, equals the ratio of
the number of active strands that separate from the network per unit
time to the current number of active strands.
Applying this definition to active strands that merged with the network
during the interval $[\tau,\tau+d\tau ]$
and separate from temporary junctions within the interval $[t,t+dt]$, 
we find that
\begin{equation}
\frac{\partial^{2} n_{\rm a}}{\partial t\partial \tau}(t,\tau,\omega)=-
\Gamma(\epsilon(t),\omega)
\frac{\partial n_{\rm a}}{\partial \tau}(t,\tau,\omega).
\end{equation}
Changes in the function $n_{\rm a}(t,0,\omega)$ are governed
by two processes at the micro-level: 
(i) detachment of active strands from temporary nodes,
and (ii) transition of passive meso-domains into the active state
under loading.
The kinetic equation for this function reads
\begin{equation}
\frac{\partial n_{\rm a}}{\partial t}(t,0,\omega)=-
\Gamma(\epsilon(t),\omega) n_{\rm a}(t,0,\omega)
+\frac{\partial N_{\rm a}}{\partial \epsilon}(\epsilon(t),\omega)
\frac{d\epsilon}{dt}(t).
\end{equation}
The solution of Eq. (7) with initial condition (4) is given by
\begin{eqnarray}
n_{\rm a}(t,0,\omega) &=& N_{\rm a}(0,\omega)\exp \biggl [ -
\int_{0}^{t} \Gamma(\epsilon(s),\omega)ds \biggr ]
\nonumber\\
&&+\int_{0}^{t} \frac{\partial N_{\rm a}}{\partial \epsilon}
(\epsilon(\tau),\omega) \frac{d\epsilon}{dt}(\tau)
\exp \biggl [ - \int_{\tau}^{t} \Gamma(\epsilon(s),\omega)ds \biggr ]
d\tau .
\end{eqnarray}
It follows from Eq. (6) that 
\begin{equation}
\frac{\partial n_{\rm a}}{\partial \tau}(t,\tau,\omega)
=\varphi(\tau,\omega)
\exp \biggl [ - \int_{\tau}^{t} \Gamma(\epsilon(s),\omega)ds \biggr ],
\end{equation}
where
\begin{equation}
\varphi(\tau,\omega)
=\frac{\partial n_{\rm a}}{\partial \tau}(t,\tau,\omega)\biggr |_{t=\tau}.
\end{equation}
To determine the function $\varphi(t,\omega)$, we use the identity
\begin{equation} 
n_{\rm a}(t,t,\omega)=n_{\rm a}(t,0,\omega)
+\int_{0}^{t} \frac{\partial n_{\rm a}}{\partial \tau}(t,\tau,\omega) d\tau.
\end{equation}
Equations (5) and (11) imply that
\begin{equation}
n_{\rm a}(t,0,\omega)
+\int_{0}^{t} \frac{\partial n_{\rm a}}{\partial \tau}(t,\tau,\omega) d\tau
=N_{\rm a}(\epsilon(t),\omega).
\end{equation}
Differentiating Eq. (12) with respect to time and using Eq. (10), we obtain
\[
\varphi(t,\omega)+\frac{\partial n_{\rm a}}{\partial t}(t,0,\omega)
+\int_{0}^{t} \frac{\partial^{2} n_{\rm a}}{\partial t\partial \tau}(t,\tau,\omega)
d\tau=\frac{\partial N_{\rm a}}{\partial \epsilon}(\epsilon(t),\omega)
\frac{d\epsilon}{dt}(t).
\]
This equality together with Eqs. (6), (7) and (11) results in 
\begin{eqnarray}
\varphi(t,\omega) &=& \Gamma(\epsilon(t),\omega)\biggl [
n_{\rm a}(t,0,\omega)
+\int_{0}^{t} \frac{\partial n_{\rm a}}{\partial \tau}(t,\tau,\omega) d\tau
\biggr ]
\nonumber\\
&=&  \Gamma(\epsilon(t),\omega)n_{\rm a}(t,t,\omega).
\end{eqnarray}
Substituting expression (13) into Eq. (9) and using Eq. (5), we
arrive at the formula
\begin{equation}
\frac{\partial n_{\rm a}}{\partial \tau}(t,\tau,\omega)
=\Gamma(\epsilon(t),\omega)N_{\rm a}(\epsilon(t),\omega)
\exp \biggl [ - \int_{\tau}^{t} \Gamma(\epsilon(s),\omega)ds \biggr ].
\end{equation}
The kinetics of rearrangement of strands in active MRs
is described by Eqs. (2), (3), (8) and (14). 
These relations are determined by (i) the distribution function
$p(\omega)$ for active MRs with various potential energies
$\omega$, (ii) the function $\Gamma_{\rm a}(\epsilon)$ that
characterizes the effect of strains on the attempt rate,
and (iii) the function 
\begin{equation}
\kappa_{\rm a}(\epsilon)=\frac{X_{\rm a}(\epsilon)}{X},
\end{equation}
that reflects mechanically-induced activation of passive MRs.

Rearrangement of strands in active MRs reflects
the viscoelastic response of a semicrystalline polymer.
The viscoplastic behavior is associated with 
the mechanically-induced slippage of junctions 
with respect to their positions in the bulk material.

Denote by $\epsilon_{\rm u}(t)$ the average strain induced by sliding
of junctions between macromolecules (the subscript index ``u" means
that $\epsilon_{\rm u}(t)$ is associated with the residual strain in a 
specimen which is suddenly unloaded at instant $t$).
The elastic strain (that reflects elongation of active strands  in 
a network) is denoted by $\epsilon_{\rm e}(t)$.
The functions $\epsilon_{\rm e}(t)$ and $\epsilon_{\rm u}(t)$ are 
connected with the macro-strain $\epsilon(t)$ by
the conventional formula
\begin{equation}
\epsilon(t)=\epsilon_{\rm e}(t)+\epsilon_{\rm u}(t).
\end{equation}
We adopt the first order kinetics for slippage of junctions with
respect to the bulk material, which implies that the increment of the
viscoplastic strain, $d\epsilon_{\rm u}$, induced by the growth of 
the macro-strain, $\epsilon$, by an increment, $d\epsilon$, 
is proportional to the absolute value of the stress $\sigma$,
\begin{equation}
\frac{d\epsilon_{\rm u}}{d\epsilon}=B |\sigma |\; {\rm sign} 
\Bigl ( \sigma\frac{d\epsilon}{dt}\Bigr ),
\end{equation}
where the pre-factor $B$ is a non-negative function of stress, 
strain  and the strain rate,
\[ B=B\Bigl (\sigma, \epsilon, \frac{d\epsilon}{dt}\Bigr ). \]
The last multiplier in Eq. (17) determines the direction of 
the viscoplastic flow of junctions.
It is convenient to present Eq. (17) in the form
\begin{equation}
\frac{d\epsilon_{\rm u}}{dt}(t)=B\Bigl (\sigma(t), \epsilon(t),
\frac{d\epsilon}{dt}\Bigr ) | \sigma(t) |\; 
{\rm sign} \Bigl [ \sigma(t) \frac{d\epsilon}{dt}(t) \Bigr ]
\frac{d\epsilon}{dt}(t),
\qquad
\epsilon_{\rm u}(0)=0.
\end{equation}
It follows from Eq. (18) that the rate of sliding vanishes when
the strain is constant.

\section{The strain energy density}

Any strand is modelled as a linear elastic solid with the mechanical energy
\[
w(t)=\frac{1}{2}\mu e^{2}(t), 
\]
where $\mu$ is the average rigidity per strand
and $e$ is the strain from the stress-free state to the deformed state.

For strands belonging to passive meso-domains, the strain $e$
coincides with $\epsilon_{\rm e}$.
Multiplying the strain energy per strand by the number of strands in
passive MRs, we find the mechanical energy of meso-domains 
where rearrangement of chains is prevented by surrounding lamellae,
\begin{equation}
W_{\rm p}(t)=\frac{1}{2}\mu X_{\rm p}(\epsilon(t))\epsilon_{\rm e}^{2}(t). 
\end{equation}
With reference to the conventional theory of temporary networks \cite{TE92},
we assume that stresses in dangling strands totally relax before
these strands merge with the network.
This implies that the reference (stress-free) state of a strand that
merges with the network at time $\tau$ coincides with
the deformed state of the network at that instant.
For active strands that have not been rearranged until time $t$,
the strain $e(t)$ coincides with $\epsilon_{\rm e}(t)$, 
whereas for active strands that have last been merged with the network 
at time $\tau\in [0,t]$, the strain $e(t,\tau)$ is given by
\[ 
e(t,\tau)=\epsilon_{\rm e}(t)-\epsilon_{\rm e}(\tau).
\]
Summing the mechanical energies of active strands 
belonging to active MRs with various potential energies, $\omega$,
that were rearranged at various instants, $\tau\in [0,t]$, we find the
mechanical energy of active meso-domains,
\begin{equation}
W_{\rm a}(t) = \frac{1}{2}\mu \int_{0}^{\infty} d\omega 
 \biggl \{ n_{\rm a}(t,0,\omega)\epsilon_{\rm e}^{2}(t)
+\int_{0}^{t} \frac{\partial n_{\rm a}}{\partial \tau}(t,\tau,\omega)
\Bigl [ \epsilon_{\rm e}(t)-\epsilon_{\rm e}(\tau)\Bigr ]^{2} d\tau \biggr \}.
\end{equation}
The mechanical energy per unit mass of a polymer reads
\[ W(t)=W_{\rm a}(t)+W_{\rm p}(t). \]
Substituting expressions (19) and (20) into this equality and using Eq. (16),
we arrive at the formula
\begin{eqnarray}
W(t) &=&  \frac{1}{2}\mu \biggl \{  X_{\rm p}(\epsilon(t))
\Bigl (\epsilon(t)-\epsilon_{\rm u}(t) \Bigr )^{2}(t)
+  \int_{0}^{\infty} d\omega  \biggl [ n_{\rm a}(t,0,\omega)
\Bigl (\epsilon(t)-\epsilon_{\rm u}(t) \Bigr )^{2}
\nonumber\\
&& +\int_{0}^{t} \frac{\partial n_{\rm a}}{\partial \tau}(t,\tau,\omega)
\Bigl ( \Bigl ( \epsilon(t)-\epsilon_{\rm u}(t)\Bigr )
-\Bigl ( \epsilon(\tau)-\epsilon_{\rm u}(\tau)\Bigr ) \Bigr )^{2} d\tau \biggr ]
\biggr \}.
\end{eqnarray}
Differentiation of Eq. (21) with respect to time results in
\begin{equation}
\frac{dW}{dt}(t)=\mu A(t)\biggl [ \frac{d\epsilon}{dt}(t)
-\frac{d\epsilon_{\rm u}}{dt}(t)\biggr ]+\frac{1}{2}\mu A_{0}(t),
\end{equation}
where
\begin{eqnarray}
A(t) &=& X_{\rm p}(\epsilon(t))\Bigl [ \epsilon(t)-\epsilon_{\rm u}(t)\Bigr ]
+\int_{0}^{\infty} d\omega \biggl \{ n_{\rm a}(t,0,\omega)
\Bigl [ \epsilon(t)-\epsilon_{\rm u}(t)\Bigr ]
\nonumber\\
&& +\int_{0}^{t} \frac{\partial n_{\rm a}}{\partial \tau}(t,\tau,\omega)
\Bigl [ \Bigl (\epsilon(t)-\epsilon_{\rm u}(t)\Bigr )
-\Bigl (\epsilon(\tau)-\epsilon_{\rm u}(\tau)\Bigr )\Bigr ]d\tau \biggr \},
\nonumber\\
A_{0}(t) &=& \frac{\partial X_{\rm p}}{\partial \epsilon}(\epsilon(t))
\frac{d\epsilon}{dt}(t) \Bigl [ \epsilon(t)-\epsilon_{\rm u}(t)\Bigr ]^{2}
+\int_{0}^{\infty} d\omega \biggl \{ \frac{\partial n_{\rm a}}{\partial t}
(t,0,\omega)\Bigl [ \epsilon(t)-\epsilon_{\rm u}(t)\Bigr ]^{2}
\nonumber\\
&& +\int_{0}^{t} \frac{\partial^{2} n_{\rm a}}{\partial t\partial \tau}(t,\tau,\omega)
\Bigl [ \Bigl (\epsilon(t)-\epsilon_{\rm u}(t)\Bigr )
-\Bigl (\epsilon(\tau)-\epsilon_{\rm u}(\tau)\Bigr )\Bigr ]^{2} d\tau \biggr \}.
\end{eqnarray}
Bearing in mind Eqs. (5) and (11), we transform the first equality in Eq. (23) 
as follows:
\begin{eqnarray*}
A(t) &=& \Bigl [ X_{\rm p}(\epsilon(t))+\int_{0}^{\infty} N_{\rm a}(\epsilon(t),\omega)
d\omega \Bigr ] \Bigl [ \epsilon(t)-\epsilon_{\rm u}(t)\Bigr ]
\nonumber\\
&& -\int_{0}^{\infty} d\omega 
\int_{0}^{t} \frac{\partial n_{\rm a}}{\partial \tau}(t,\tau,\omega)
\Bigl [\epsilon(\tau)-\epsilon_{\rm u}(\tau) \Bigr ]d\tau .
\end{eqnarray*}
This formula together with Eqs. (1) and (3) implies that
\begin{equation}
A(t) = X \Bigl [ \epsilon(t)-\epsilon_{\rm u}(t)\Bigr ]
-\int_{0}^{\infty} d\omega 
\int_{0}^{t} \frac{\partial n_{\rm a}}{\partial \tau}(t,\tau,\omega)
\Bigl [\epsilon(\tau)-\epsilon_{\rm u}(\tau) \Bigr ]d\tau .
\end{equation}
Substitution of expressions (6) and (7) into the second equality in Eq. (23)
yields
\begin{equation}
A_{0}(t)=\Bigl [ \frac{\partial X_{\rm p}}{\partial \epsilon}(\epsilon(t))
+\int_{0}^{\infty} \frac{\partial N_{\rm a}}{\partial \epsilon}(\epsilon(t),\omega)
d\omega \Bigr ]
\frac{d\epsilon}{dt}(t) \Bigl [ \epsilon(t)-\epsilon_{\rm u}(t)\Bigr ]^{2}
-A_{1}(t),
\end{equation}
where
\begin{eqnarray}
A_{1}(t) &=& \int_{0}^{\infty} \Gamma(\epsilon(t),\omega) d\omega
\biggl \{ n_{\rm a}(t,0,\omega) \Bigl [ \epsilon(t)-\epsilon_{\rm u}(t)\Bigr ]^{2}
\nonumber\\
&& +\int_{0}^{t} \frac{\partial n_{\rm a}}{\partial \tau}(t,\tau,\omega)
\Bigl [ \Bigl (\epsilon(t)-\epsilon_{\rm u}(t)\Bigr )
-\Bigl (\epsilon(\tau)-\epsilon_{\rm u}(\tau)\Bigr )\Bigr ]^{2} d\tau \biggr \}.
\end{eqnarray}
It follows from Eqs. (1), (3) and (25) that
\[ A_{0}(t)=-A_{1}(t). \]
This equality together with Eq. (22) results in
\begin{equation}
\frac{dW}{dt}(t)=\mu \Bigl [ A(t)  \frac{d\epsilon}{dt}(t)
-\frac{1}{2}\Bigl (A_{1}(t)+A_{2}(t)\Bigr )\Bigr ],
\end{equation}
where
\begin{equation}
A_{2}(t)=2A(t)\frac{d\epsilon_{\rm u}}{dt}(t).
\end{equation}

\section{Constitutive equations}

For uniaxial loading with small strains at the reference 
temperature $T_{0}$, the Clausius-Duhem inequality reads
\cite{Hau00}
\begin{equation}
T_{0}\frac{dQ}{dt}(t)=-\frac{dW}{dt}(t)+\frac{1}{\rho}\sigma(t)\frac{d\epsilon}{dt}(t)
\geq 0,
\end{equation}
where $\rho$ is mass density,
and $Q$ is the entropy production per unit mass.
Substitution of expression (27) into Eq. (29) implies that
\begin{equation}
T_{0}\frac{dQ}{dt}(t)=\frac{1}{\rho}\Bigl [ \sigma(t)-\rho\mu A(t)\Bigr ] \frac{d\epsilon}{dt}(t)
+\frac{1}{2}\Bigl [ A_{1}(t)+A_{2}(t)\Bigr ] \geq 0.
\end{equation}
Because Eq. (30) is to be fulfilled for an arbitrary program 
of straining, $\epsilon=\epsilon(t)$, the expression in the first square 
brackets vanishes.
This assertion together with Eq. (24) results in the stress--strain relation
\begin{eqnarray}
\sigma(t) &=& \rho\mu A(t)
\nonumber\\
&=& E \biggl \{ \Bigl [ \epsilon(t)-\epsilon_{\rm u}(t)\Bigr ]
-\frac{1}{X}\int_{0}^{\infty} d\omega 
 \int_{0}^{t} \frac{\partial n_{\rm a}}{\partial \tau}(t,\tau,\omega)
\Bigl [\epsilon(\tau)-\epsilon_{\rm u}(\tau) \Bigr ]d\tau \biggr \},
\end{eqnarray}
where
\[ E=\rho\mu X \]
is an analog of the Young modulus.
It follows from Eqs. (18), (28) and (31) that
\begin{equation}
A_{2}(t)=\frac{2}{\rho\mu} B\Bigl (\sigma(t), \epsilon(t),\frac{d\epsilon}{dt}(t)\Bigr )
\sigma^{2}(t)\Bigl | \frac{d\epsilon}{dt}(t)\Bigr |.
\end{equation}
According to Eqs. (26) and (32), the functions $A_{1}(t)$ and $A_{2}(t)$
are non-negative for an arbitrary program of loading,
which, together with Eq. (31), implies that the Clausius--Duhem
inequality (30) is satisfied.

Substitution of Eqs. (3), (14) and (15) into Eq. (31) results in the constitutive 
equation
\begin{eqnarray}
\sigma(t) &=& E \biggl \{ \Bigl [ \epsilon(t)-\epsilon_{\rm u}(t)\Bigr ]
-\kappa_{\rm a}(\epsilon(t)) \int_{0}^{\infty} p(\omega) d\omega 
\nonumber\\
&&\times \int_{0}^{t} \Gamma(\epsilon(t),\omega) 
\exp \biggl [-\int_{\tau}^{t} \Gamma(\epsilon(s),\omega) ds\biggr ]
\Bigl [\epsilon(\tau)-\epsilon_{\rm u}(\tau) \Bigr ]d\tau \biggr \}.
\end{eqnarray}
Given functions $p(\omega)$, $\Gamma_{\rm a}(\epsilon)$ and $\kappa_{\rm a}(\epsilon)$,
the time-dependent response of a semicrystalline polymer 
at isothermal uniaxial loading with small strains is
determined by Eqs. (2), (18) and (33).
For a standard relaxation test with the longitudinal strain $\epsilon^{0}$,
\[
\epsilon(t)=\left \{\begin{array}{ll}
0,  & t<0,
\\
\epsilon^{0}, & t\geq 0,
\end{array}\right .
\]
these equations imply that
\[
\sigma(t,\epsilon^{0})=E(\epsilon^{0}-\epsilon_{\rm u}^{0})\biggl \{ 1-\kappa_{\rm a}(\epsilon^{0})
\int_{0}^{\infty} p(\omega) \biggl [ 1-\exp\Bigl (-\Gamma_{\rm a}(\epsilon^{0})
\exp(-\omega)t\Bigr )\biggr ] d\omega\biggr \},
\]
where $\epsilon_{\rm u}^{0}$ is the strain induced by sliding of junctions.
Introducing the notation
\begin{equation}
C_{1}(\epsilon^{0})=E(\epsilon^{0}-\epsilon_{\rm u}^{0}),
\qquad
C_{2}(\epsilon^{0})=E(\epsilon^{0}-\epsilon_{\rm u}^{0})
\kappa_{\rm a}(\epsilon^{0}),
\end{equation}
we present this equality as follows:
\begin{equation}
\sigma(t,\epsilon^{0})=C_{1}(\epsilon^{0})-C_{2}(\epsilon^{0})
\int_{0}^{\infty} p(\omega) \biggl [ 1-\exp\Bigl (-\Gamma_{\rm a}(\epsilon^{0})
\exp(-\omega)t\Bigr )\biggr ] d\omega.
\end{equation}
To fit experimental data, we adopt the random energy model \cite{Dyr95} with
\begin{equation}
p(\omega)=p_{0}\exp \biggl [-\frac{(\omega-\Omega)^{2}}{2\Sigma^{2}}\biggr ]
\quad (\omega\geq 0),
\qquad
p(\omega)=0
\quad (\omega <0),
\end{equation}
where $\Omega$ and $\Sigma$ are adjustable parameters,
and the pre-factor $p_{0}$ is determined by the condition
\begin{equation}
\int_{0}^{\infty} p(\omega)d\omega =1.
\end{equation}
Given a strain $\epsilon^{0}$, Eqs. (35) and (36) are determined by
5 material constants:
\begin{enumerate}
\item 
the average potential energy for rearrangement of strands $\Omega$,

\item
the standard deviation of distribution of potential energies
$\Sigma$,

\item
the attempt rate for separation of strands in active MRs from
their junctions $\Gamma_{\rm a}$,

\item
the coefficients $C_{1}$ and $C_{2}$.
\end{enumerate}
Our purpose now is to find these parameters by fitting experimental data.

\section{Fitting of observations}

We begin with matching the relaxation curve measured 
at the strain $\epsilon_{8}=0.025$.
Because the rate of rearrangement, $\Gamma_{\rm a}$, and the average
activation energy, $\Omega$, are mutually dependent 
[Eqs. (35) and (36) imply that the growth of $\Omega$ results in an increase 
in $\Gamma_{\rm a}$], we set $\Gamma_{\rm a}=1$ s and approximate 
the relaxation curve by using 4 experimental constants: 
$\Omega$, $\Sigma$, $C_{1}$ and $C_{2}$.
To find these quantities, we fix the intervals 
$[0,\Omega_{\max}]$ and $[0,\Sigma_{\max}]$, 
where the ``best-fit" parameters $\Omega$ and $\Sigma$ are assumed 
to be located, and divide these intervals into $J$ subintervals by
the points 
$\Omega_{i}=i\Delta_{\Omega}$ 
and $\Sigma_{j}=j\Delta_{\Sigma}$  ($i,j=1,\ldots,J$)
with $\Delta_{\Omega}=\Omega_{\max}/J$,
$\Delta_{\Sigma}=\Sigma_{\max}/J$.
For any pair, $\{\Omega_{i},\Sigma_{j} \}$, the integral in Eq. (35) 
is evaluated numerically (by Simpson's method with 200 points and the step
$\Delta_{\omega}=0.1$).
The pre-factor $p_{0}$ is determined by Eq. (37).
The coefficients $C_{1}=C_{1}(i,j)$ and $C_{2}=C_{2}(i,j)$ that
minimize the function
\begin{equation}
{\cal J}(i,j)=\sum_{t_{m}} \Bigl [ \sigma_{\rm exp}(t_{m})-\sigma_{\rm num}(t_{m}) \Bigr ]^{2},
\end{equation}
where the sum is calculated over all experimental points $t_{m}$,
are found by the least-squares method.
The stress $\sigma_{\rm exp}$ in Eq. (38) is measured in the relaxation
test, whereas the quantity $\sigma_{\rm num}$ is given by Eq. (35).
The ``best-fit" parameters $\Omega$ and $\Sigma$ minimize the function ${\cal J}$ 
on the set
\[
\Bigl \{ \Omega_{i},\; \Sigma_{j} \quad (i,j=1,\ldots, J) \Bigr \}.
\]
After determining the  ``best-fit" constants, $\Omega_{i}$ and $\Sigma_{j}$, 
this procedure is repeated for the new intervals $[\Omega_{i-1},\Omega_{i+1}]$
and $[\Sigma_{j-1},\Sigma_{j+1}]$ to ensure good accuracy of fitting.
Figure 2 demonstrates fair agreement between the experimental data 
and the results of numerical simulation with $\Omega=5.03$ and $\Sigma=3.05$.

To approximate relaxation curves at other strains, $\epsilon_{k}$,
we fix the constants $\Omega$ and $\Sigma$ found by matching experimental
data at $\epsilon_{8}$ and fit every relaxation curve by using 
3 adjustable parameters: $\Gamma_{\rm a}$, $C_{1}$ and $C_{2}$.
To find these quantities, an algorithm is applied similar
to that employed to match the relaxation curve at $\epsilon_{8}=0.025$.
We fix the interval $[0,\Gamma_{\max}]$, where the ``best-fit" attempt rate
$\Gamma_{\rm a}$ is supposed to be located,
and divide this interval into $J$ subintervals by the points 
$\Gamma_{i}=i\Delta_{\Gamma}$  ($i=1,\ldots,J$)
with $\Delta_{\Gamma}=\Gamma_{\max}/J$.
For any $\Gamma_{i}$, we calculate the integral in
Eq. (35) numerically and find the coefficients $C_{1}=C_{1}(i)$ 
and $C_{2}=C_{2}(i)$ minimizing the function (38)
by the least-squares method.
The ``best-fit" attempt rate is determined from the
condition of minimum for the function ${\cal J}$ 
on the set
$\Bigl \{ \Gamma_{i} \quad (i=1,\ldots, J) \Bigr \}$.
After finding the ``best-fit"  value, $\Gamma_{i}$, 
this procedure is repeated for the new interval $[\Gamma_{i-1},\Gamma_{i+1}]$
to ensure an acceptable accuracy of fitting.
Figures 2 to 4 show good agreement between the observations
and the results of numerical analysis.

For any longitudinal strain $\epsilon_{k}$, the attempt rate, $\Gamma_{\rm a}(\epsilon_{k})$,
is determined by matching an appropriate relaxation curve.
The fraction of active MRs, $\kappa_{\rm a}(\epsilon_{k})$, is found from
Eq. (34),
\[ 
\kappa_{\rm a}(\epsilon_{k})=\frac{C_{2}(\epsilon_{k})}{C_{1}(\epsilon_{k})}.
\]
These quantities are plotted versus strain $\epsilon$ in Figures 5 and 6.
The experimental data are approximated by the phenomenological equations
\begin{equation}
\log \Gamma_{\rm a}=\gamma_{0}+\gamma_{1}\epsilon,
\qquad
\kappa_{\rm a}=k_{0}+k_{1}\epsilon,
\end{equation}
where the coefficients $\gamma_{i}$ and $k_{i}$ are found by the
least-squares method.
Figures 5 and 6 reveal two different regimes of
the nonlinear viscoelastic behavior of polypropylene.
In the region of relatively small strains (less than $\epsilon_{\ast}=0.02$), 
the quantities $\log \Gamma_{\rm a}$ and $\kappa_{\rm a}$
increase linearly with strain, $\epsilon$.
In the region of strains exceeding the threshold value, $\epsilon_{\ast}$,
the attempt rate grows rather weakly, 
and the fraction of active MRs remains constant.
This observation implies that the free-volume theory 
may be applied to the description of the nonlinear viscoelastic 
response of polypropylene at relatively small strains only.
When strains exceed the threshold value, $\epsilon_{\ast}$, 
relaxation of longitudinal stresses becomes strain-independent.

This conclusion results in two questions of interest: 
\begin{enumerate}
\item
What is a micro-mechanism for transition from the nonlinear 
(strain-dependent) to the linear (strain-independent)
viscoelastic behavior of isotactic polypropylene at the
threshold strain, $\epsilon_{\ast}$.

\item
Whether other semicrystalline polymers demonstrate  the nonlinear 
time-dependent response in a limited range of strains only,
i.e., whether relaxation of stresses in semicrystalline polymers 
(unlike amorphous glassy polymers) becomes
strain-independent at threshold strains, $\epsilon_{\ast}$, that
are noticeably lower than the yield strain, $\epsilon_{\rm y}$.
\end{enumerate}
These issues will be the subject of a subsequent study.
 
\section{Concluding remarks}

Constitutive equations have been derived for the time-dependent behavior
of semicrystalline polymers at isothermal loading with small strains.
To develop stress--strain relations, a version of the mean-field approach
is employed: a complicated micro-structure of isotactic polypropylene
is replaced by an equivalent transient network of macromolecules bridged by
junctions (physical cross-links, entanglements and crystalline lamellae).
The network is assumed to be strongly inhomogeneous, and it is
thought of as an ensemble of meso-regions with various potential energies 
for separation of strands from temporary nodes.

The viscoelastic response of a semicrystalline polymer is ascribed 
to the processes of detachment and reformation of chains in active 
meso-domains.
Rearrangement of active strands  is modelled as a thermo-mechanically 
activated process whose rate obeys the Eyring formula.

The viscoplastic response is described by slippage of junctions 
with respect to their positions in the bulk material.
The rate of sliding of junctions is assumed to be proportional to 
the macro-stress in a specimen.

The mechanical energy is determined as the sum of the strain energies 
of active strands.
Constitutive equations are derived by using the laws of thermodynamics.
These relations are applied to study relaxation of longitudinal stresses
at uniaxial tension of specimens.

A series of relaxation tests have been performed on isotactic
polypropylene at room temperature.
Adjustable parameters in the stress--strain equations are found
by fitting observations.
The following conclusions are drawn from the analysis of experimental data:
\begin{enumerate}
\item
Despite a complicated morphology of isotactic polypropylene,
its time-dependent response is rheologically simple in the sense
that the relaxation spectrum (which is determined by distribution
function, $p(\omega)$, for potential energies of active MRs)
is not affected by mechanical factors.

\item
Two regimes of the nonlinear viscoelastic behavior of polypropylene
are observed.
At relatively small strains (less than the threshold
value $\epsilon_{\ast}=0.02$), the attempt rate, $\Gamma_{\rm a}$, 
and the fraction of active MRs, $\kappa_{\rm a}$, increase with strain, 
$\epsilon$, in agreement with the free-volume concept.
When the strain, $\epsilon$, exceeds its threshold value,
$\epsilon_{\ast}$, the attempt rate, $\Gamma_{\rm a}$, 
grows with strain rather weakly, while the fraction of active MRs, 
$\kappa_{\rm a}$, remains constant.
\end{enumerate}

\setlength{\unitlength}{0.8 mm}
\begin{figure}[tbh]
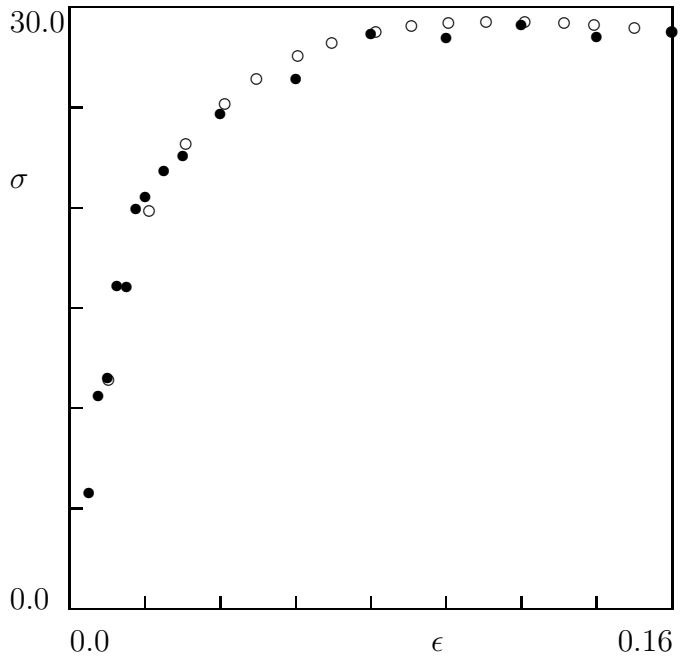

\begin{center}

\end{center}
\vspace*{5 mm}

\caption{\small The longitudinal stress $\sigma$ MPa versus 
strain $\epsilon$.
Unfilled circles: experimental data in a tensile test 
with the cross-head speed 5 mm/min.
Filled circles: experimental data at the beginnings of
relaxation tests}
\end{figure}

\begin{figure}[tbh]
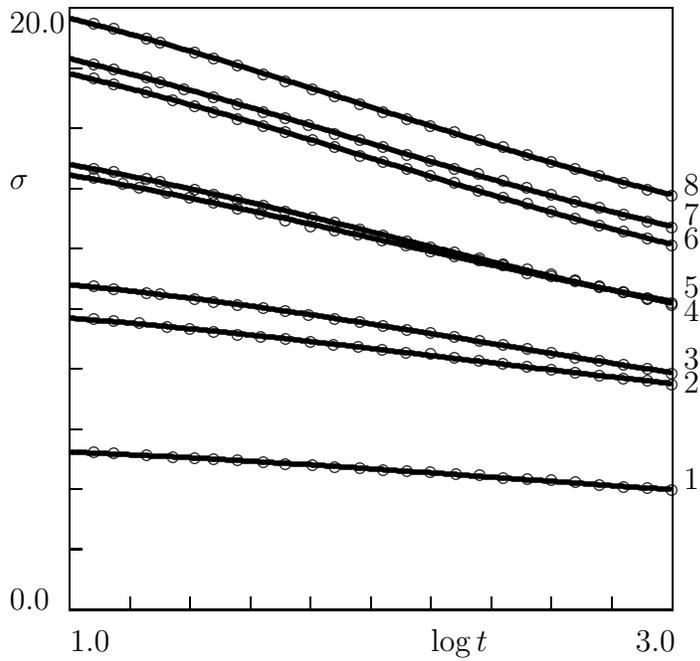

\begin{center}
%
\end{center}
\vspace*{5 mm}

\caption{\small The longitudinal stress $\sigma$ MPa
versus time $t$ s in a tensile relaxation test 
at a strain $\epsilon$.
Circles: experimental data.
Solid lines: results of numerical simulation.
Curve 1: $\epsilon=0.005$;
curve 2: $\epsilon=0.0075$;
curve 3: $\epsilon=0.010$;
curve 4: $\epsilon=0.0125$;
curve 5: $\epsilon=0.015$;
curve 6: $\epsilon=0.0175$;
curve 7: $\epsilon=0.020$;
curve 8: $\epsilon=0.025$}
\end{figure}

\begin{figure}[tbh]
\begin{center}
%
\end{center}
\vspace*{5 mm}

\caption{\small The longitudinal stress $\sigma$ MPa
versus time $t$ s in a tensile relaxation test 
at a strain $\epsilon$.
Circles: experimental data.
Solid lines: results of numerical simulation.
Curve 1: $\epsilon=0.03$;
curve 2: $\epsilon=0.04$;
curve 3: $\epsilon=0.06$;
curve 4: $\epsilon=0.08$}
\end{figure}

\begin{figure}[tbh]
\begin{center}
%
\end{center}
\vspace*{5 mm}

\caption{\small The longitudinal stress $\sigma$ MPa
versus time $t$ s in a tensile relaxation test 
at a strain $\epsilon$.
Circles: experimental data.
Solid lines: results of numerical simulation.
Curve 1: $\epsilon=0.10$;
curve 2: $\epsilon=0.12$;
curve 3: $\epsilon=0.14$;
curve 4: $\epsilon=0.16$}
\end{figure}

\begin{figure}[t]
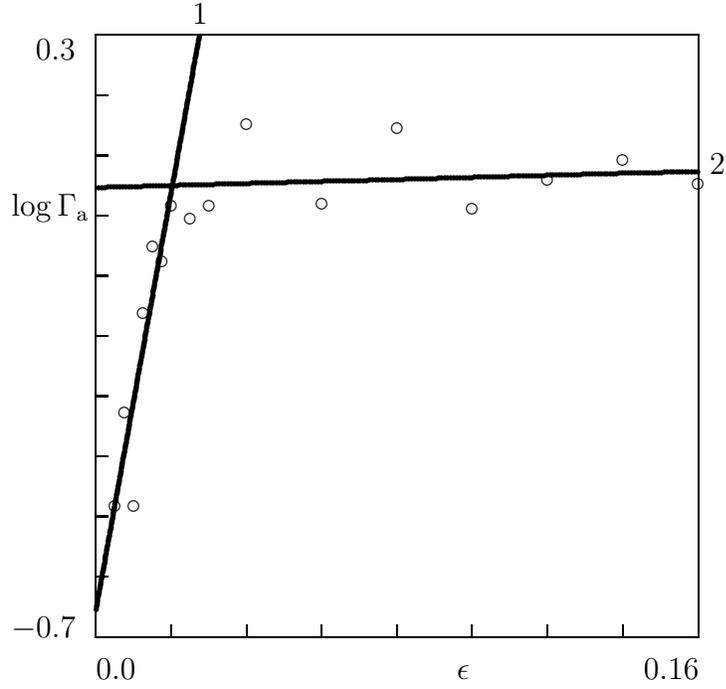

\begin{center}
%
\end{center}
\vspace*{5 mm}

\caption{\small The attempt rate $\Gamma_{\rm a}$ s$^{-1}$
versus strain $\epsilon$ in tensile relaxation tests.
Circles: treatment of observations.
Solid lines: approximation of the experimental data by Eq. (39).
Curve 1: $\gamma_{0}=-0.66$, $\gamma_{1}=34.73$;
curve 2: $\gamma_{0}=0.05$, $\gamma_{1}=0.17$}
\end{figure}

\begin{figure}[tbh]
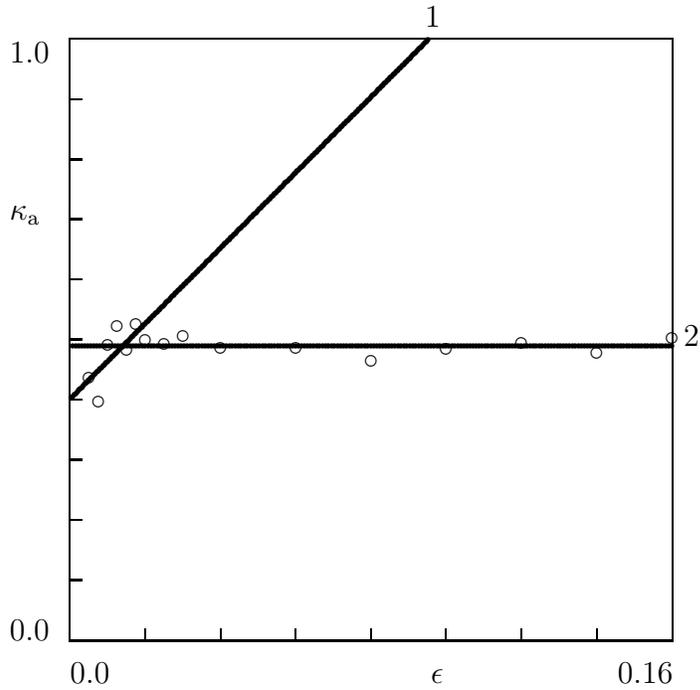

\begin{center}
%
\end{center}
\vspace*{5 mm}

\caption{\small The fraction of active MRs $\kappa_{\rm a}$
versus strain $\epsilon$ in tensile relaxation tests.
Circles: treatment of observations.
Solid lines: approximation of the experimental data by Eq. (39).
Curve 1: $k_{0}=0.40$, $k_{1}=6.28$;
curve 2: $k_{0}=0.49$, $k_{1}=0.0$}
\end{figure}

\end{document}